\documentclass[twocolumn,showpacs,pre,superscriptaddress]{revtex4}

\usepackage{graphicx}
\usepackage{amssymb}

\begin{document}

\title{Feedback-optimized parallel tempering Monte Carlo}

\author{Helmut G.~Katzgraber}
\affiliation{Theoretische Physik, ETH Z\"urich, 
CH-8093 Z\"urich, Switzerland}

\author{Simon Trebst}
\affiliation{Theoretische Physik, ETH Z\"urich, 
CH-8093 Z\"urich, Switzerland}
\affiliation{Computational Laboratory,  ETH Zentrum, 
CH-8092 Z\"urich, Switzerland}
\affiliation{Microsoft Research and Kavli Institute for Theoretical Physics, 
University of California, Santa Barbara, CA 93106, USA}

\author{David A.~Huse}
\affiliation{Department of Physics, Princeton University, 
Princeton, NJ 08544, USA}

\author{Matthias Troyer}
\affiliation{Theoretische Physik, ETH Z\"urich,
CH-8093 Z\"urich, Switzerland}

\date{\today}

\begin{abstract}
We introduce an algorithm to systematically improve the efficiency of 
parallel tempering Monte Carlo simulations by optimizing the simulated 
temperature set. Our approach is closely related to a recently introduced 
adaptive algorithm that optimizes the simulated statistical ensemble in 
generalized broad-histogram Monte Carlo simulations. 
Conventionally, a temperature set is chosen in 
such a way that the acceptance rates for replica swaps between adjacent 
temperatures are independent of the temperature and large enough to ensure 
frequent swaps. In this paper, we show that by choosing the temperatures 
with a modified version of the optimized ensemble feedback method we can 
minimize the round-trip times between the lowest and highest temperatures 
which effectively increases the efficiency of the parallel tempering 
algorithm.  In particular, the density of temperatures in the optimized 
temperature set increases at the ``bottlenecks'' of the simulation, such 
as phase transitions. In turn, the acceptance rates are now temperature 
dependent in the optimized temperature ensemble.
We illustrate the feedback-optimized parallel tempering algorithm 
by studying the two-dimensional Ising ferromagnet and the two-dimensional 
fully-frustrated Ising model, and briefly discuss possible feedback schemes
for systems that require configurational averages, such as spin glasses.
\end{abstract}

\maketitle

\section{Introduction}
\label{sec:introduction}

The free energy landscapes of complex systems are characterized by many 
local minima that are separated by entropic barriers. The simulation of 
such systems with conventional Monte Carlo methods is slowed down by long 
relaxation times due to the suppressed tunneling through these barriers. 
Extended ensemble simulations address this problem by broadening the range 
of phase space which is sampled in the respective reaction coordinate. 
Recently, an adaptive algorithm \cite{trebst:04} has been introduced that 
explores entropic barriers by sampling a broad histogram in a reaction 
coordinate and iteratively optimizes the simulated statistical ensemble 
defined in the reaction coordinate to speed up equilibration. The key idea 
of the approach is to measure the local diffusivity along the reaction 
coordinate, thereby identifying the bottlenecks in the simulations and then 
using this information to systematically shift statistical weights towards 
these bottlenecks in a feedback loop. The optimized histogram converges to 
a characteristic form exhibiting peaks at the bottlenecks of the 
simulation, e.g., in the vicinity of the entropic barriers. The simulation 
of an optimized ensemble results in equilibration times that can be 
substantially lower compared to other extended ensemble simulations that 
aim at sampling a flat histogram in the respective reaction coordinate 
\cite{trebst:04,dayal:04}. Flat-histogram algorithms include the 
multicanonical method \cite{berg:91,berg:92}, broad histograms 
\cite{deoliveira:96} and transition matrix Monte Carlo \cite{wang:02} when 
combined with entropic sampling, as well as the adaptive algorithm of Wang 
and Landau \cite{wang:01,wang:01a}.

Parallel tempering Monte Carlo \cite{hukushima:96} has proven to be 
a strong ``workhorse'' in fields as diverse as chemistry, physics, 
biology, engineering, and material science \cite{earl:05}. Similar to 
replica Monte Carlo \cite{swendsen:86}, simulated tempering
\cite{marinari:92}, or extended ensemble methods \cite{lyubartsev:92}, 
the algorithm aims to overcome entropic barriers in 
the free energy landscape by simulating a broad range of temperatures. 
This allows the system to escape metastable states when wandering to 
higher temperatures and subsequently relaxing at lower temperatures again 
on time scales that are substantially smaller than conventional 
simulations at a fixed temperature. In this paper, we maximize the 
efficiency of parallel tempering Monte Carlo by optimizing the 
distribution of temperature points in the simulated temperature set such 
that round-trip rates of replicas between the two extremal temperatures in 
the simulated temperature set are maximized. The optimized temperature 
sets are determined by an iterative feedback algorithm that is closely 
related to the previously mentioned adaptive ensemble optimization method 
for broad-histogram Monte Carlo simulations \cite{trebst:04}. The feedback 
method concentrates temperature points near the bottleneck of a 
simulation, typically in the vicinity of a phase transition or the ground 
state of the system. As a consequence, we find that for the optimal choice 
of temperatures the acceptance probabilities for swap moves between 
neighboring temperature points show a strong modulation with temperature 
and are not independent of temperature as suggested in several recent 
approaches 
\cite{kofke:02,kofke:04,rathore:05,predescu:04,predescu:05,kone:05}.

The paper is structured as follows: In Sec.~\ref{sec:pt} we present a 
detailed introduction of the parallel tempering Monte Carlo method. In 
Sec.~\ref{sec:ensembe_optimization} we generalize the feedback method of 
Ref.~\onlinecite{trebst:04} to the parallel tempering Monte Carlo algorithm. 
Results on two paradigmatic models, the two-dimensional Ising ferromagnet and 
the two-dimensional fully-frustrated Ising model are presented in 
Sec.~\ref{sec:results}, as well as a discussion on how to proceed with systems
that require configurational averages, such as spin glasses. Concluding 
remarks follow in Sec.~\ref{sec:conclusions}.

\section{Parallel tempering Monte Carlo}
\label{sec:pt}

In the parallel tempering Monte Carlo algorithm \cite{swendsen:86,marinari:92, 
lyubartsev:92,hukushima:96}, $M$ non-interacting replicas of the system are 
simulated simultaneously at a range of temperatures $\{T_1, T_2, \ldots, 
T_M\}$. After a fixed number of Monte Carlo sweeps a sequence of swap 
moves, the exchange of two replicas at neighboring temperatures, $T_i$ and 
$T_{i+1}$, is suggested and accepted with a probability
\begin{equation}
  p(E_i, T_i \rightarrow E_{i+1}, T_{i+1}) 
  = \min \left\{ 1,  \exp(\Delta\beta \Delta E)  \right\}~,
\end{equation}
where $\Delta\beta = 1/T_{i+1} - 1/T_i$ is the difference between the 
inverse temperatures and $\Delta E = E_{i+1} -E_i$ is the difference in 
energy of the two replicas. At a given temperature, an accepted swap move 
effects in a global update as the current configuration of the system is 
exchanged with a replica from a nearby temperature. For a given replica, 
the swap moves induce a random walk in temperature space. This random walk 
allows the replica to overcome free energy barriers by wandering to high 
temperatures where equilibration is rapid and return to low temperatures 
where relaxation times can be long. The simulated system can thereby 
efficiently explore complex energy landscapes that can be found in 
frustrated spin systems \cite{diep:05}, spin glasses 
\cite{binder:86,mezard:87,young:98} or proteins \cite{wales:03}. While the 
simulation of $M$ replicas takes $M$ times more CPU time, the speedup 
attained with parallel tempering Monte Carlo can be orders of magnitude 
larger. 
In addition, one often wishes to measure observables as a function 
of temperature. Thus parallel tempering Monte 
Carlo delivers already several measurements at different temperatures in 
one simulation. In order to maximize the number of statistically 
independent visits at low temperatures, we want to maximize for each 
replica the number of round-trips between the lowest and highest 
temperature, $T_1$ and $T_M$, respectively. The rate of round-trips of a 
replica strongly depends on the simulated statistical ensemble, that is 
the choice of temperature points $\{T_1, T_2, \ldots, T_M\}$ in the 
parallel tempering simulation. 

In this paper, we present an algorithm that 
systematically optimizes the simulated temperature set to maximize the 
number of round-trips between the two extremal temperatures $T_1$ and 
$T_M$ for each replica and thereby substantially improve equilibration of 
the system at all temperatures. Conventional approaches assume that to 
achieve this goal, the simulated temperature set should be chosen in such a 
way that the probability for replica swap moves at neighboring 
temperatures should be ``flat'', that is approximately independent of 
temperature. If the specific heat of the system is assumed to be constant, 
then a good approximation for such a temperature set can be attained with 
a geometric progression \cite{predescu:04}. 
Given a temperature range $[T_1, T_M]$, the intermediate $M-2$ temperatures 
can be computed via
\begin{equation}
T_k = T_1\prod_{i = 1}^{k - 1} R_i \;\;\;\;\;\;\;\;\;\;\;\;
 R_i = \sqrt[M-1]{\frac{T_M}{T_1}} \; .
\label{eq:geometric}
\end{equation}
The geometric progression peaks the number of temperatures around the 
minimum temperature $T_1$ where a slower relaxation is assumed. This is 
not optimal when the system has a diverging specific heat (at an intermediate
temperature): Because in 
order to ensure enough overlap between the energy distributions of 
neighboring temperatures $\Delta T_{i,i+1} \sim C_{V}T_i/\sqrt{N}$, where 
$C_V$ is the specific heat per spin and $N$ the number of spins, the 
acceptance probabilities are inversely correlated to the functional 
behavior of $C_V$ via the inverse beta function law \cite{predescu:04}. 
Thus, for example at a phase transition where the specific heat diverges, 
the acceptance probabilities for a geometric temperature set will show a 
pronounced dip (cf.~Sec.~\ref{sec:FM}). More complex methods such as the 
approach by Kofke \cite{kofke:02,kofke:04}, its improvement by Rathore 
{\em et al.}~\cite{rathore:05}, as well as the method suggested by 
Predescu \cite{predescu:04,predescu:05} aim to obtain acceptance 
probabilities for the parallel tempering moves that are independent of 
temperature by compensating for the effects of the specific heat. In 
particular, Kone and Kofke \cite{kone:05} suggest that an acceptance 
probability of 23\% is optimal. In this work we show that this is not 
necessarily the optimal case.

\section{Temperature set optimization}
\label{sec:ensembe_optimization}

\begin{figure}
\includegraphics[width=0.95\columnwidth]{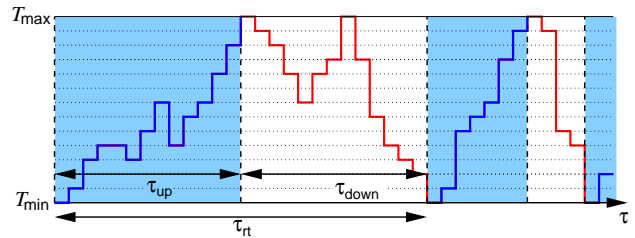}
\caption{(Color online) 
Sketch of the random walk that a given replica performs in temperature 
space in the course of the simulation. Ideally, the replica will wander up 
($\tau_{\rm up}$) and down ($\tau_{\rm down}$) in the simulated temperature 
range $[T_{\rm min}, T_{\rm max}]$. The goal of the feedback method is to 
maximize the number of round trips each replica performs in this 
temperature range, and thereby minimize the average round-trip time 
$\tau_{\rm rt} = \tau_{\rm up} + \tau_{\rm down}$.
}
\label{fig:times}
\end{figure}

Our goal is to vary the temperature set $\{ T_i\}$ of a parallel tempering 
simulation in such a way that for a given system we speed up 
equilibration at all temperatures. To accomplish this, we maximize 
the rate of round trips that each replica performs between the two 
extremal temperatures $T_{\rm min} = T_1$ and $T_{\rm max} = T_M$ 
following a similar approach to the ensemble optimization technique 
presented in Ref.~\cite{trebst:04}. 
For a given temperature set, we can measure the diffusion of a replica 
through temperature space by adding a label ``up'' or ``down'' to the 
replica that indicates which of the two extremal temperatures, 
$T_{\rm min}$ or $T_{\rm max}$ respectively, the replica has visited most 
recently. The label of a replica changes only when the replica visits the 
opposite extreme. For instance, the label of a replica with label ``up'' 
remains unchanged if the replica returns to the lowest temperature 
$T_{\rm min}$, but changes to ``down'' 
upon its first visit to $T_{\rm max}$. This is illustrated in 
Fig.~\ref{fig:times}. For each temperature point in the temperature set 
$\{ T_i\}$, we record two histograms $n_{\rm up}(T_i)$ and $n_{\rm 
down}(T_i)$. Before attempting a sequence of swap moves, we increment that 
histogram at temperature $T_i$ which has the label of the respective 
replica currently at temperature $T_i$. If a replica has not yet visited 
either of the two extremal temperatures, we increment neither of the 
histograms. This allows us to evaluate for each temperature point the
fraction of replicas which have visited one of the two extremal 
temperatures most recently (e.g., $T_{\rm min}$) as
\begin{equation}
f(T_i) =  \frac{n_{\rm up}(T_i)}{n_{\rm up}(T_i) + n_{\rm down}(T_i)} \; .
\label{fraction}
\end{equation}

The labeled replicas define a steady-state current $j$ from $T_{\rm min}$ 
to $T_{\rm max}$ that is independent of temperature, e.g., the rate at which
``up'' walkers are created at $T_{\rm min}$ and -- in equilibrium -- absorbed
at $T_{\rm max}$. 
In the following we assume that $T$ is a continuous variable, independent
of the temperature points in the current temperature set. We can then determine
the current $j$ to first order in the derivative as
\begin{equation}
   j = D(T) \eta(T) \frac{df}{dT} \;,
   \label{eq:current}
\end{equation}  
where $D(T)$ is the local diffusivity at temperature $T$ and the derivative 
$df/dT$ is estimated by a linear regression based on the measurements in 
Eq.~(\ref{fraction});  $\eta (T)$ is a density distribution indicating 
the probability for a replica to reside at temperature $T$. 
We approximate $\eta (T)$ with a step-function 
$\eta(T) = C /\Delta T$, where $\Delta T = T_{i+1}-T_i$ is the length of 
the temperature interval around temperature $T_i < T < T_{i+1}$ for the 
current temperature set. The normalization constant $C$ is chosen such 
that
\begin{equation}
  \int_{T_1}^{T_M} \eta(T) dT = C \int_{T_1}^{T_M} \frac{dT}{\Delta T} = 1 \;.
  \label{eq:normalization}
\end{equation}
Rearranging Eq.~(\ref{eq:current}) gives a simple measure of the local 
diffusivity $D(T)$ of a replica at temperature $T$
\begin{equation}
  D(T) \sim \frac{\Delta T}{df/dT} \;,
  \label{eq:diffusivity}
\end{equation}
where we have dropped the normalization constant $C$ and the current $j$
which is constant for any specific choice of temperature set.

To increase the efficiency of the algorithm, we maximize the 
current $j$ in temperature space by varying the simulated temperature set 
$\{ T_i \}$ and thus varying the probability distribution $\eta (T)$ 
between the two extremal temperatures, $T_{\rm min}$ and $T_{\rm max}$, 
which are not  changed. In Ref.~\onlinecite{trebst:04} it has been shown that 
the optimal probability distribution $\eta^{\rm opt} (T)$ is inversely 
proportional to the square root of the local diffusivity $D(T)$:
\begin{equation}
  \eta^{\rm opt} (T)  \propto  \frac{1}{\sqrt{D(T)}}  \;.
\end{equation}
For the optimal distribution of temperature points the fraction $f^{\rm 
opt}(T)$ then decays as
\begin{equation}
  \frac{df}{dT}^{\rm opt} = \eta^{\rm opt} (T) \propto \frac{1}{\Delta T^{\rm opt}} \;,
\end{equation}
which implies that for any given temperature interval $\Delta T = T_{i+1} 
- T_i$ of the optimal temperature set the fraction has a {\em constant} 
decay
\begin{equation}
  \Delta f^{\rm opt} = f^{\rm opt}(T_{i}) - f^{\rm opt}(T_{i+1}) = 1/(M-1) \;,
  \label{eq:linearfraction}
\end{equation} 
where $M$ is the number of replicas. 

In our algorithm we approach the optimal temperature set and its 
respective probability distribution by iteratively feeding back 
measurements of the local diffusivity. After measuring the diffusion of 
replicas for a given temperature set an improved probability distribution 
$\eta'(T)$ is found as
\begin{equation}
  \eta'(T) = \frac{C'}{\Delta T'} = C' \sqrt{ \frac{1}{\Delta T} \, 
  \frac{df}{dT} }  \;,
\end{equation}
where the normalization constant $C'$ is again chosen so that the 
normalization condition in Eq.~(\ref{eq:normalization}) is met. The 
step-function $\eta'(T)$ is still defined for the {\em original} 
temperature set, that is the jumps in the function occur at the 
temperature points in $\{ T_i \}$. The {\em optimized} temperature set $\{ 
T'_i\}$ is then found by choosing the $k$-th temperature point $T'_k$ such 
that
\begin{equation}
  \int_{T'_1}^{T'_k} \eta'(T) dT = \frac{k}{M} \;,
  \label{eq:feedback}
\end{equation}
where $1<k<M$ and the two extremal temperatures $T'_1=T_1$ and $T'_M=T_M$ 
remain fixed.

We summarize the feedback algorithm by the following sequence of steps
\begin{itemize}
\item Start with a trial temperature set $\{ T_i \}$.
\item Repeat
\begin{itemize}
\item[$\circ$] Reset the histograms $n_{\rm up}(T) = n_{\rm down}(T) = 0$.
\item[$\circ$] For the current temperature set perform a parallel tempering simulation with $N_{\rm sw}$ swap moves.
          After each sequence of swap moves:
\begin{itemize}
\item[] Update the labels of all replicas.
\item[] Record histograms $n_{\rm up}(T)$ and $n_{\rm down}(T)$.
\end{itemize}
\item[$\circ$] For the given temperature set estimate an optimized probability distribution $\eta'(T)$ via 
          \[\eta'(T)  = C' \sqrt{ \frac{1}{\Delta T} \, \frac{df}{dT} } \;.\]
\item[$\circ$] Obtain the optimized temperatures $\{ T'_i \}$ via
         \[\int_{T_1}^{T'_k} \eta'(T) dT = \frac{k}{M} \;. \]
\item[$\circ$] Increase the number of swaps $N_{\rm sw} \leftarrow 2N_{\rm sw}$.
\end{itemize}
\item Stop once the temperature set $\{ T_i \}$ has converged.
\end{itemize}
The initial number of swaps $N_{\rm sw}$ should be chosen large enough 
such that a few of round-trips are recorded, thereby ensuring that 
steady-state data for $n_{\rm up}(T) $ and $n_{\rm down}(T)$ are measured. 
The derivative $df/dT$ can be determined by a linear regression, where the 
number of regression points is flexible.  Initial batches with the limited 
statistics of only a few round trips may require a larger number of 
regression points than subsequent batches with smaller round-trip times 
and better statistics.

\section{Results}
\label{sec:results}

\subsection{Ferromagnetic Ising model}
\label{sec:FM}

In order to illustrate the feedback method, we start with a simple test 
model, the two-dimensional ferromagnetic Ising model (FM). The Hamiltonian 
for the model is given by
\begin{equation}
{\mathcal H}_{\rm FM} = - J\sum_{\langle i,j\rangle} S_i S_j \; ,
\label{eq:ham_fm}
\end{equation}
where $J = 1$ and $S_i = \pm 1$ represent Ising spins on a square lattice 
with $N = L^2$ spins. In our simulations we apply periodic boundary 
conditions and consider nearest-neighbor interactions only. The simple 
Ising model with uniform couplings has no 
frustration or disorder, and exhibits a second-order phase transition 
at $T_{\rm c} = 2/\ln(1+\sqrt{2}) \approx 2.269$ from a magnetically 
ordered to a paramagnetic phase.

\begin{figure}
\includegraphics[width=0.95\columnwidth]{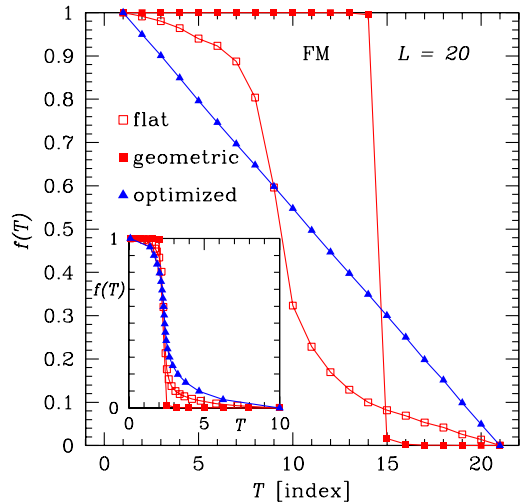}
\vspace*{-1.0cm}

\caption{(Color online)
Fraction $f(T)$ of replicas diffusing from the lowest to the highest 
temperature as a function of the temperature index for the ferromagnetic 
Ising model. For the initial temperature set based on a geometric 
progression (filled squares), the fraction shows a sharp drop between two 
temperature points. A similar behavior is found for a temperature set where
the acceptance rates are constant $\sim 40\%$ independent of temperature 
(temperature set with ``flat'' acceptance rates,  open squares). 
In contrast, for the optimized temperature set (triangles) the fraction 
constantly decreases. The inset shows the fraction as a 
function of temperature $T$. The dashed line in the inset represents the 
critical temperature of the two-dimensional Ising model, 
$T_{\rm c} \approx 2.269$. 
}
\label{fig:fm_f_L20}
\end{figure}

For simplicity, we define an initial temperature set $\{ T_i \}$ with 
$M=21$ temperature points performing a geometric progression, 
Eq.~(\ref{eq:geometric}), for a temperature interval $[T_1 = 0.1, T_M = 
10.0]$. The minimum temperature $T_1$ is chosen low enough such that the system 
can approach the zero-temperature ground state of the model, and the 
maximum temperature $T_M$ is chosen well above the critical region of the 
phase transition. For a short parallel tempering simulation ($N_{\rm sw} = 
1.6 \cdot 10^7$, one parallel tempering swap after each lattice sweep) using 
this initial temperature set, we measure the diffusive current of replicas 
wandering from the lowest to the highest temperature using an additional 
label for each replica as described above. In Fig.~\ref{fig:fm_f_L20} we 
show the fraction of replicas diffusing ``up" in temperature space. 
For the geometric temperature set a
sharp drop between two temperature points is observed close to the 
critical region of the phase transition at $T_c\approx 2.269$. 
Similar results are also found for a ``flat'' temperature set where the
acceptance rates are approximately constant around $\sim 40 \%$ 
(with fluctuations of $\sim 10\%$) and independent of temperature, although 
the drop is not as pronounced. 

\begin{figure}
\includegraphics[width=0.95\columnwidth]{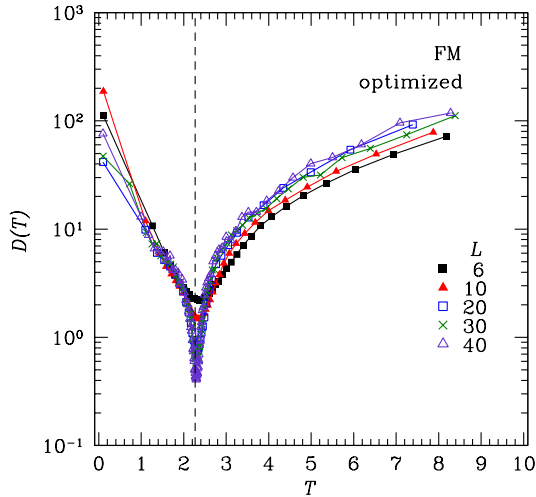}
\vspace*{-1.0cm}
                                                                                
\caption{(Color online)
Local diffusivity $D(T)$ of the random walk a replica performs in temperature space 
for the ferromagnetic Ising model as a function of temperature $T$ after the 
feedback optimization for several system sizes $L$. Notice the logarithmic 
vertical scale. The vertical dashed line represents $T_{\rm c} \approx 2.269$.
}
\label{fig:fm_diff_opt}
\end{figure}

Calculating the local diffusivity $D(T)$ for the random walk in temperature 
space using Eq.~(\ref{eq:diffusivity}), we find a strong suppression around 
this 
critical region as illustrated in Fig.~\ref{fig:fm_diff_opt}. When increasing 
the size of the simulated system, the dip in the local diffusivity further 
proliferates, an additional indicator that the slowdown of the random walk 
in temperature space is dominated by the occurrence of a phase transition. 
Note the logarithmic scale of the ordinate axis in Fig.~\ref{fig:fm_diff_opt}.

\begin{figure}
\includegraphics[width=0.95\columnwidth]{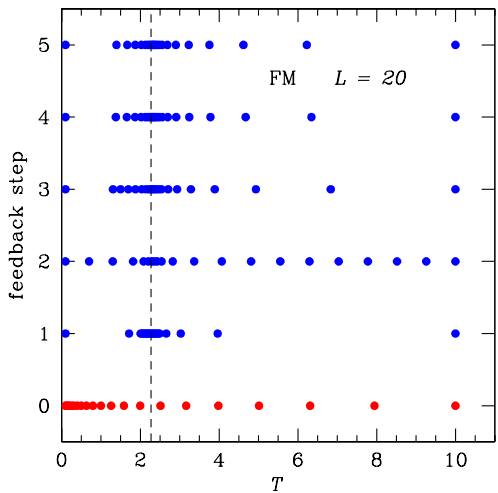}
\vspace*{-1.0cm}
                                                                                
\caption{(Color online)
Temperature sets for the ferromagnetic Ising model for different feedback 
steps. Starting from a geometric progression temperature set (step 0), we 
apply a feedback loop until the temperature set converges. While the 
geometric progression places many temperatures at low temperatures, the 
density of temperatures after the feedback optimization is highest at the 
bottleneck of the simulation around the critical temperature (marked by a 
vertical dashed line). Rapid convergence of the optimized temperature set 
is found after 3 -- 4 feedback steps and a total of $N_{\rm sw} \approx 
1.6 \cdot 10^7$ swap moves in our parallel tempering simulations.
}
\label{fig:fm_t-sets_L20}
\end{figure}

When applying the feedback, Eq.~(\ref{eq:feedback}), to define a new 
temperature set, this suppression in the local diffusivity leads to a 
concentration of temperature points near the critical temperature as shown 
in Fig.~\ref{fig:fm_t-sets_L20}. The feedback thereby tries to compensate 
for the reduced mobility of replicas in the critical region by 
reallocating resources towards this temperature range. In contrast, the 
density of temperatures close to the lowest temperature is greatly 
reduced, thereby suppressing constant swapping of replicas at low 
temperatures where for the initial temperature set multiple replicas 
converged to the fully-polarized ground state configuration. 

\begin{figure}
\includegraphics[width=0.95\columnwidth]{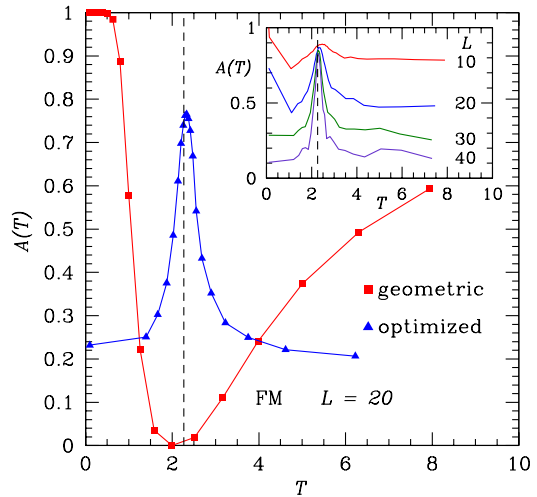}
\vspace*{-1.0cm}

\caption{(Color online) 
Acceptance probabilities $A(T)$ as a function of temperature $T$ for the 
ferromagnetic Ising model. The inset shows the 
acceptance rates as a function of temperature in the optimized case for 
varying system sizes $L$ and a fixed number of temperature points. The
vertical dashed line marks the critical temperature.
}
\label{fig:fm_a_L20}
\end{figure}

This effect 
becomes even more evident when measuring the acceptance probabilities for 
swap moves as illustrated in Fig.~\ref{fig:fm_a_L20}. The acceptance 
probabilities for the initial temperature set based on a geometric 
progression saturate close to unity for temperatures below $T \lesssim 
0.7$, whereas they show a pronounced dip already for small system sizes 
($L = 20$) around the critical temperature (marked by a vertical dashed 
line).  In contrast, the feedback-optimized temperature set shows a 
pronounced peak in the acceptance rate $A(T)$ near the critical 
temperature where temperature points are accumulated by the feedback. Away 
from the critical temperature region the acceptance probabilities drop.
The inset of Fig.~\ref{fig:fm_a_L20} shows the acceptance probabilities 
$A(T)$ for the optimized temperature sets for a {\em fixed} number of replicas 
and varying sizes of the simulated system. While the acceptance probability 
around the critical temperature remains nearly constant, the exact value 
away from the critical region decreases with increasing system size. 
This ``mean" acceptance probability away from the bottleneck of 
the simulation can be tuned by varying the number of simulated replicas.

The fact that for the optimized temperature set the acceptance 
probabilities vary with temperature contradicts various alternative 
approaches in the literature 
\cite{kofke:04,predescu:04,kone:05,rathore:05,predescu:05} that aim at 
choosing a temperature set in such a way that the acceptance probability 
of attempted swaps becomes independent of temperature. Naively, one might 
assume that the choice of constant acceptance rates produces an unbiased 
random walk in temperature space. This assumption is similar to the idea 
underlying generalized-ensemble algorithms that aim to sample a flat 
histogram in the energy such as the multicanonical method 
\cite{berg:91,berg:92} or the Wang-Landau algorithm 
\cite{wang:01,wang:01a}. For these flat-histogram algorithms, it has been shown 
that they cannot reproduce the scaling behavior of an unbiased Markovian 
random walk in energy space, but experience critical slowing down 
\cite{dayal:04}. This slowdown can be overcome by optimizing the simulated 
statistical ensemble by a similar feedback algorithm \cite{trebst:04} as 
presented here and sampling an optimized histogram that in general is 
{\em not} flat, but reallocates resources towards the bottleneck of the 
simulation, e.g., in the vicinity of a phase transition or close to the 
ground state of the system.

\begin{figure}
\includegraphics[width=0.95\columnwidth]{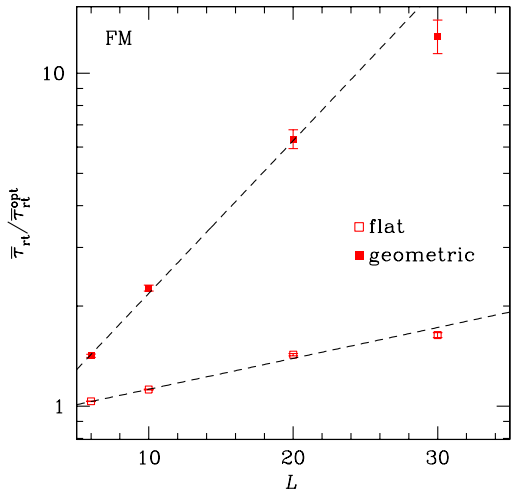}
\vspace*{-1.0cm}

\caption{(Color online) 
Average round-trip times $\overline{\tau}_{\rm rt}$ 
before the optimization divided by the average round-trip times after 
the feedback optimization 
($\overline{\tau}_{\rm rt}^{\rm opt}$) as a function of system size.
The data for the filled squares are for a system starting from a geometric
progression and represent the speedup obtained by the feedback method.
The open symbols correspond to a temperature set which initially has
``flat'' acceptance probabilities.
The dashed lines are guides to the eye.
}
\label{fig:fm_trt}
\end{figure}

Measuring the diffusion of replicas in a subsequent simulation for the 
optimized temperature set, we find that the current of replicas wandering 
from the lowest to the highest temperature is now characterized by a 
constantly decreasing fraction $f(T)$ in agreement with 
Eq.~(\ref{eq:linearfraction}) as shown in Fig.~\ref{fig:fm_f_L20} (triangles). 
In addition, we find that for the optimized temperature set 
the replicas wander evenly up and down in temperature space, that is 
$\tau_{\rm up} \approx \tau_{\rm down}$. 

In Fig.~\ref{fig:fm_trt} we show the ratio between the mean round-trip times
$\overline{\tau}_{\rm rt}$ before optimization from a geometric 
and ``flat'' temperature set divided by the mean round-trip times 
$\overline{\tau}_{\rm rt}^{\rm opt}$  after optimization in order to illustrate
the speedup in replica diffusion attained by the feedback procedure.
The data show clearly for all system sizes that the 
round-trip times after the optimization of the temperature set do not 
increase as fast as for a geometric progression  or ``flat'' temperature set. 
Note that for these temperature sets with a fixed number of temperature points
the average round-trip times before and after the feedback optimization scale
$\sim \exp(aL^b)$ for the system sizes studied.
It is important to note that our algorithm identifies the bottleneck of the 
parallel tempering simulation that in this model occurs in the form of 
critical slowing down at the phase transition solely based on measurements 
of the local diffusivity. The feedback then allows to shift additional 
resources towards this bottleneck in a quantitative way. 
In the next section, we apply the algorithm to a more complex model 
with frustration and a different type of phase transition at zero 
temperature.  

\subsection{Fully-frustrated Ising model}
\label{sec:ff}

The Hamiltonian of the fully-frustrated (FF) Ising model is given by
\begin{equation}
{\mathcal H}_{\rm FF} = -\sum_{\langle i,j\rangle} J_{ij} S_i S_j \; ,
\label{eq:ham_ff}
\end{equation}
where the spins lie on the vertices of a two-dimensional square lattice 
with periodic boundary conditions. The bonds $J_{ij}$ are chosen such that 
$|J_{ij}| = 1$, but with the constraint that the product of the bonds 
around {\em all} plaquettes of the system is negative, i.e.,
\begin{equation}
\prod_{\Box} J_{ij} = -1 \; .
\label{eq:ff_prod}
\end{equation}
The model does not order at finite temperatures, but exhibits a critical 
point at zero temperature. In the vicinity of this transition to a highly 
degenerate ground-state manifold, the system relaxes very slowly. 

\begin{figure}
\includegraphics[width=0.95\columnwidth]{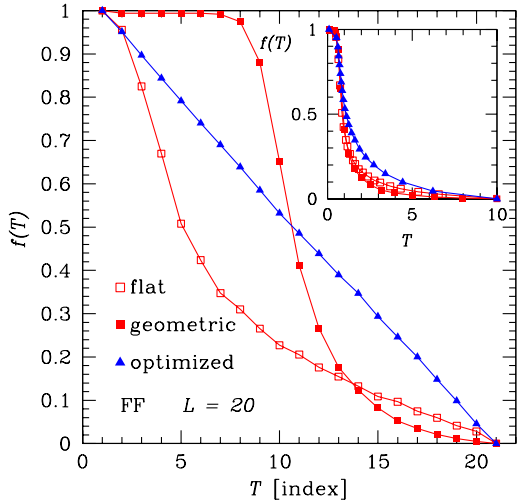}
\vspace*{-1.0cm}
                                                                                
\caption{(Color online)
Fraction $f(T)$ of replicas diffusing from the lowest to the highest 
temperature for the fully frustrated Ising model. 
Displayed are data for an initial ``flat'' temperature set 
with $M=21$ temperature points that yields temperature-independent acceptance 
probabilities for swap moves (open squares). In addition, data for a geometric
progression ($M = 21$) are also shown (filled squares). After the
optimization, the change in the fraction is independent of the temperature 
index (triangles). The inset shows the fractions as a function of 
temperature $T$. Data for $N_{\rm sw} = 2\cdot 10^6$.
}
\label{fig:ff_f_L20-variable}
\end{figure}

In this section, we discuss our feedback optimization algorithm for two
choices of the initial temperature set. First we start from the temperature set 
introduced in Sec.~\ref{sec:FM} computed with a geometric progression, 
Eq.~(\ref{eq:geometric}), which has $T_{\rm min} = 0.1$, $T_{\rm max} = 
10.0$ and $M = 21$ temperatures. In this first approach, we keep the number 
of temperature points constant for all system sizes $L$. In the second 
approach, we choose an initial temperature set where we fix again 
$T_{\rm min}$ and $T_{\rm max}$ to the aforementioned values but tune the 
number of temperatures $M$ as well as 
their position in such a way that we obtain acceptance probabilities for 
swap moves that are approximately independent of the temperature 
(``flat'') with a mean value of $A(T) \sim 40 \%$ and deviations around this 
mean value of maximally 10\% \cite{ffimproblems}.

\begin{figure}
\includegraphics[width=0.95\columnwidth]{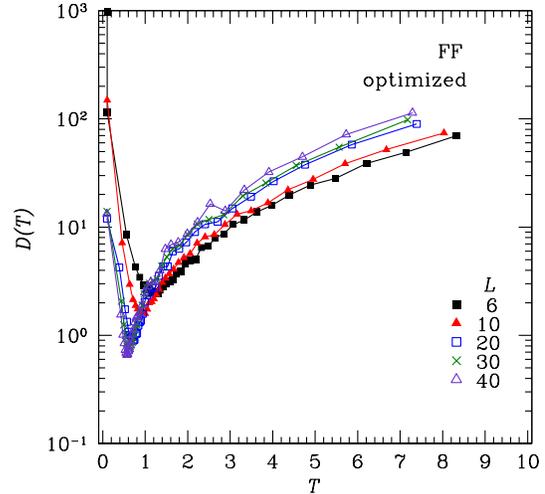}
\vspace*{-1.0cm}
                                                                                
\caption{(Color online)
Local diffusivity $D(T)$ of a random walk in temperature space for the 
fully-frustrated Ising model as a function of temperature $T$ after the 
feedback optimization for several system sizes $L$. Notice the logarithmic 
vertical scale.
}
\label{fig:ff_diff_opt}
\end{figure}

\begin{figure}
\includegraphics[width=0.95\columnwidth]{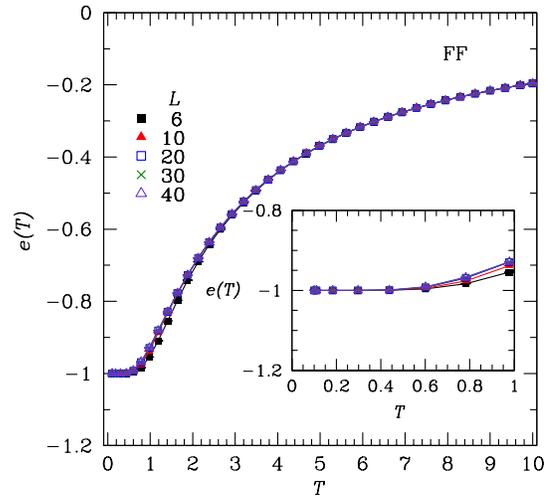}
\vspace*{-1.0cm}
                                                                                
\caption{(Color online)
Energy per spin $e(T) = (1/N)[\langle {\mathcal H}\rangle]_{\rm av}$ as a 
function of temperature $T$ for the fully-frustrated Ising model for 
several system sizes. The data show that already for $T \lesssim 0.5$ the 
energy is independent of temperature, thus signaling that the system has 
reached the ground state. The inset zooms into the temperature range 
around $T = 0$.
}
\label{fig:ff_e}
\end{figure}

\begin{figure}
\includegraphics[width=0.95\columnwidth]{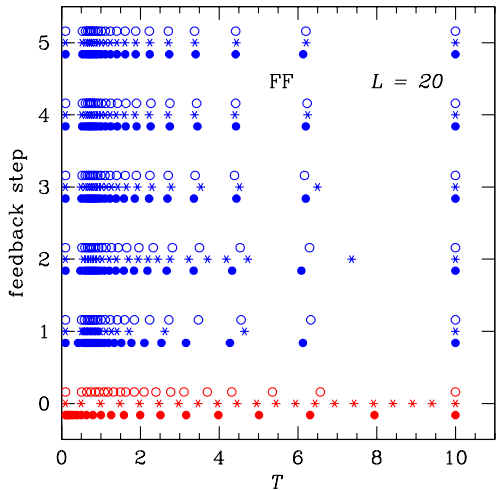}
\vspace*{-1.0cm}
                                                                                
\caption{(Color online)
Temperature sets for the fully-frustrated Ising model for different 
feedback steps. Starting from a temperature set where the acceptance 
probabilities are independent of temperature with $M =21$ temperature
points (open symbols) and a temperature set obtained by a geometric 
progression also with $M = 21$ temperature points (filled symbols), we apply 
repeated feedback steps until the temperature sets converge to the optimal 
distributions. Also shown are data for an initial temperature set with $M=21$ 
equidistant temperature points (stars). Independent of the initial 
temperature set, an optimal temperature distribution is found after 3 -- 4 
iterations and  $\sim 7.5\cdot10^6$ swaps. After the successful feedback, 
temperature points accumulate near the transition to the ground state 
slightly above zero temperature.
}
\label{fig:ff_t-sets_L20}
\end{figure}

We show the fraction $f(T)$ of replicas diffusing 
from the lowest to the highest temperature as a function of the 
temperature index in Fig.~\ref{fig:ff_f_L20-variable}. 
Similar to the Ising model, the data for the geometric progression 
temperature set deviate considerably from a straight line which is 
expected for the optimal temperature distribution. A similar behavior is found 
when starting from a temperature set that  
initially has temperature-independent acceptance probabilities. 
The local diffusivity in temperature space calculated from the measured 
diffusive current is plotted in Fig.~\ref{fig:ff_diff_opt}.  
There is a pronounced dip in the 
diffusivity around $T\approx0.5$ that we can identify as the temperature 
region where the system enters the highly degenerate ground-state 
manifold, e.g., by calculating the energy of the system, as plotted in 
Fig.~\ref{fig:ff_e}. Again we find that the diffusivity points to a strong 
bottleneck of the simulation at the critical point which for the fully 
frustrated Ising model is at the transition to the zero-temperature 
ground-state manifold. The general shape of the diffusivity in the vicinity of 
this bottleneck remains unchanged with respect to the feedback.
By applying the feedback method, additional temperature points are shifted 
towards this bottleneck which is illustrated in Fig.~\ref{fig:ff_t-sets_L20} 
for the geometric progression 
(full symbols) as well as the ``flat'' temperature set (open symbols). For 
moderately large system sizes, we again find rapid convergence of the 
generated temperature sets within 3 -- 4 feedback steps and a total of 
$N_{\rm sw} \approx 7.5\cdot10^6$ swap moves. For the optimized 
temperature set, the diffusive current is again characterized by a fraction 
of replicas drifting from the lowest to the highest temperature that  
linearly decreases with the temperature index, see 
Fig.~\ref{fig:ff_f_L20-variable} (triangles).

\begin{figure}
\includegraphics[width=0.95\columnwidth]{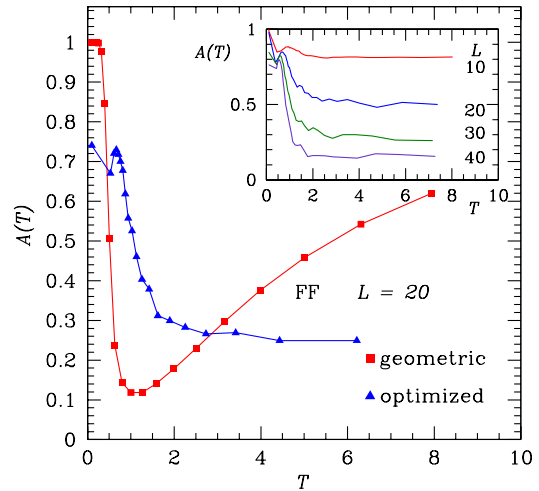}
\vspace*{-1.0cm}

\caption{(Color online) 
Acceptance probabilities $A(T)$ for replica swap moves as a function of 
temperature $T$ for the fully-frustrated Ising model. While the acceptance 
probabilities for a geometric progression temperature set show a 
pronounced dip close to $T = 0$, the optimized ensemble shows a peak close 
to zero temperature where the system enters the ground-state manifold. The 
inset shows, for a fixed number of temperatures, the acceptance 
rates as a function of temperature for different system sizes $L$.  As for 
the Ising model, the ``mean'' value away from the bottlenecks can be tuned 
by increasing the number of temperatures. This illustrates that in order 
to obtain higher acceptance rates away from the bottlenecks of the 
simulation, the number of temperatures have to be increased with increasing 
$L$.
}
\label{fig:ff_a_L20}
\end{figure}

\begin{figure}
\includegraphics[width=0.95\columnwidth]{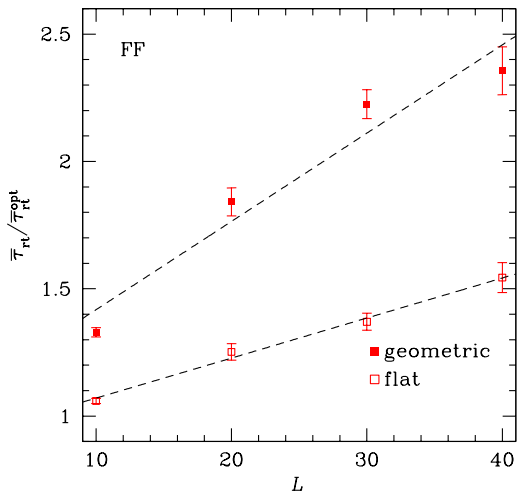}
\vspace*{-1.0cm}

\caption{(Color online) 
Average round-trip times $\overline{\tau}_{\rm rt}$ before the optimization 
divided by the average round-trip times after the feedback optimization
($\overline{\tau}_{\rm rt}^{\rm opt}$) as a function of system size for the
fully-frustrated Ising model.
The data for the filled squares are for a system starting from a geometric
progression temperature set and represent the speedup obtained by the 
feedback method. In addition, we show data for a temperature set 
with ``flat'' temperature-independent acceptance rates (open squares).
The dashed lines are guides to the eye.
}
\label{fig:ff_trt}
\end{figure}

\begin{figure}
\includegraphics[width=0.95\columnwidth]{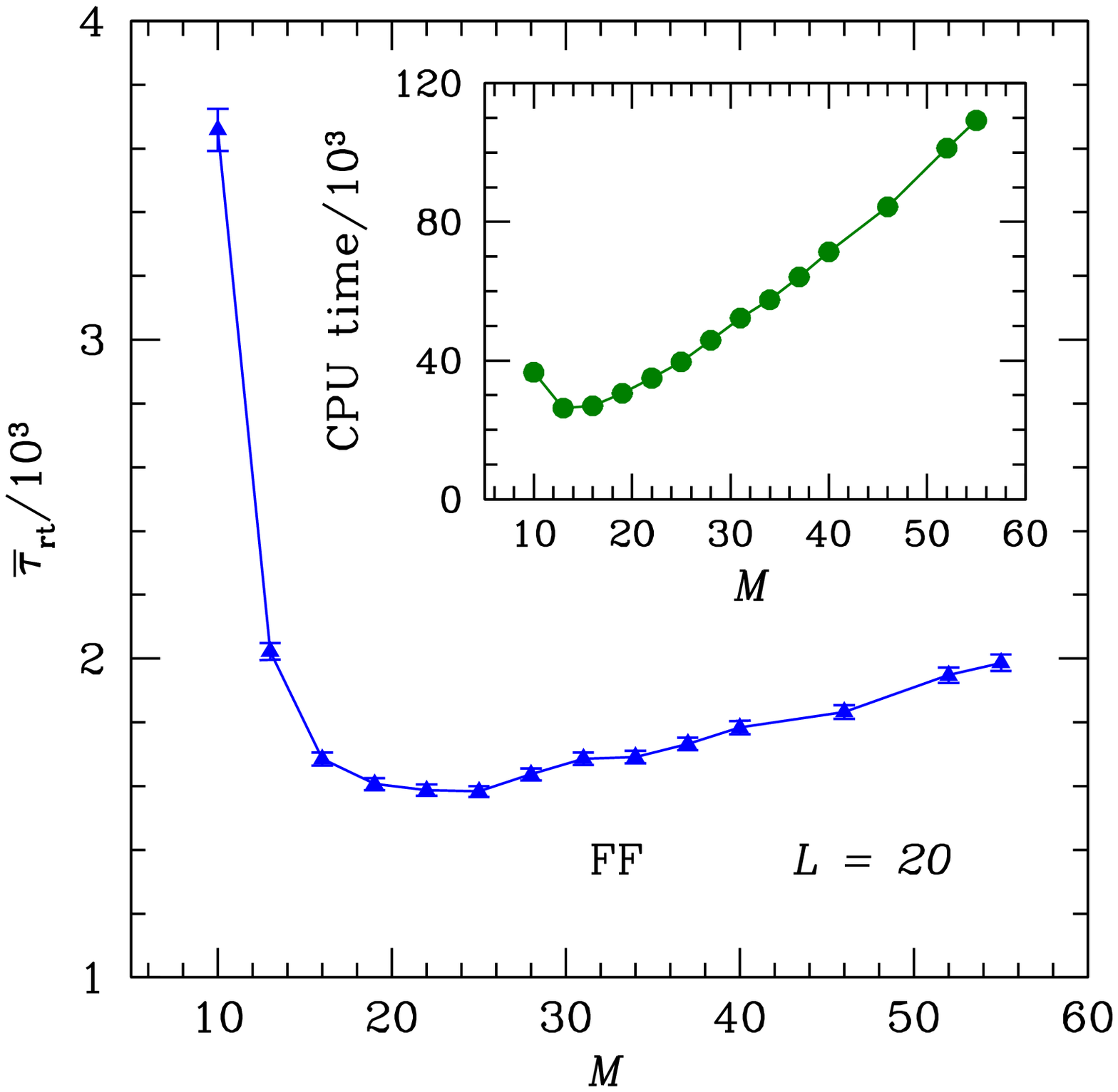}
\vspace*{-1.0cm}
                                                                                
\caption{(Color online)
Average round-trip times $\overline{\tau}_{\rm rt}$ as a function of the
number of temperatures $M$ for the fully-frustrated Ising model with 
$L = 20$ after the feedback optimization. The data show that the round-trip 
times only depend moderately on the number of temperatures $M$, provided 
that there is sufficient overlap of the energy distributions. 
For a small number of replicas,  
this is no longer the case and the round-trip times increase drastically, 
e.g., for $M \lesssim 12$ in this plot. The inset shows the CPU time which is
the product of the average round-trip time with the number of temperatures 
$M$. The data show a more pronounced minimum.
}
\label{fig:ff_m_dep}
\end{figure}

The acceptance probabilities $A(T)$ for replica swap moves are shown in 
Fig.~\ref{fig:ff_a_L20} for simulations with a geometric progression and 
the optimized temperature set. While the acceptance probabilities peak at 
unity close to zero and dramatically drop in the geometric progression 
temperature set, in the optimized set most temperatures are reshuffled to 
the low-temperature region slightly above zero temperature where the 
system enters the highly degenerate ground-state manifold. The inset to 
Fig.~\ref{fig:ff_a_L20} shows the acceptance probabilities as a function 
of temperature for a {\em fixed} number of temperatures, as well 
as $T_{\rm min} = 0.1$ and $T_{\rm max} = 10.0$ fixed. As in the case for 
the ferromagnet, the ``mean'' acceptance probability away from the ground 
state bottleneck seems almost independent of temperature and settles at a 
value that is determined by the number of temperatures used for a given 
system size $L$.

In order to test the efficiency of the feedback method on the FFIM, in 
Fig.~\ref{fig:ff_trt} we show  the ratio between the mean round-trip times
$\overline{\tau}_{\rm rt}$ before optimization divided by the mean 
round-trip times $\overline{\tau}_{\rm rt}^{\rm opt}$  after optimization in 
order to illustrate the speedup in replica diffusion attained by the feedback 
procedure. The data show clearly for all system sizes that the
round-trip times after the optimization of the temperature set do not
increase as fast as for a geometric progression or ``flat'' temperature set.
For these temperature sets where the number of temperature points 
increases with system size we find that the average round-trip times
scale $\sim a + bx^{c}$.

Finally, we discuss the effects of varying the number of temperatures $M$ 
in the temperature set. Figure 
\ref{fig:ff_m_dep} shows the average round-trip times for the fully 
frustrated Ising model ($L = 20$) as a function of the number of temperatures 
$M$. For $M \gtrsim 12$, the average
round-trip times show only moderate variations with the number of temperature
points $M$, whereas for a smaller number of temperatures the average
round-trip times increase drastically. 
This can be understood by keeping in mind that a parallel tempering swap will 
only be accepted with a high probability if the energy distributions between 
neighboring temperatures overlap. If the minimum and maximum temperatures 
are fixed and $M$ is reduced, the energy distributions will cease to overlap, 
which accounts for the increased average round-trip times. 
Factoring in the total CPU time, which we define as the product of the 
average round-trip time with the number of temperatures, the minimum is more 
pronounced (inset to Fig.~\ref{fig:ff_m_dep}) and clearly illustrates that 
while a larger number of temperatures has little effect on the round-trip 
times, the total CPU time increases drastically with increasing $M$.

Because the data for the average round-trip times vs $M$ have an optimal
value, it is conceivable to develop a feedback optimization method 
that optimizes both the position of the temperatures and the number of
temperatures $M$. Furthermore, in addition to optimizing the number and 
locations of the temperature points, we have also explored varying the 
ratio of the number of sweep moves attempted to the number of replica-swap
moves attempted, since this is yet another parameter one must set in a
parallel tempering simulation. This ratio can be adjusted globally 
(the same ratio at all temperatures) or locally (the ratio optimized 
independently at each temperature).  This will be discussed in more detail 
in a subsequent communication.

\subsection{Spin glasses}
\label{sec:sg}

Since the optimization of temperature sets improves the sampling for the 
two paradigmatic spin models discussed above, it is a natural step to ask how 
this feedback optimization technique can be applied to improve state-of-the-art 
parallel tempering simulations of Ising spin glass models, such as the 
three-dimensional (3D) Edwards-Anderson Ising spin 
glass \cite{edwards:75,binder:86}:
\begin{equation}
{\mathcal H}_{\rm SG} = -\sum_{\langle i,j\rangle} J_{ij} S_i S_j .
\label{eq:ham_sg}
\end{equation}
Here the spins lie on the vertices of a cubic lattice
with periodic boundary conditions. The bonds $J_{ij}$ are chosen according to
a Gaussian distribution with zero mean and standard deviation unity.
The system undergoes a spin-glass transition at $T_{\rm c} = 0.951(9)$ 
\cite{bhatt:88,marinari:98,katzgraber:06}.

\begin{figure}[t]
\includegraphics[width=0.95\columnwidth]{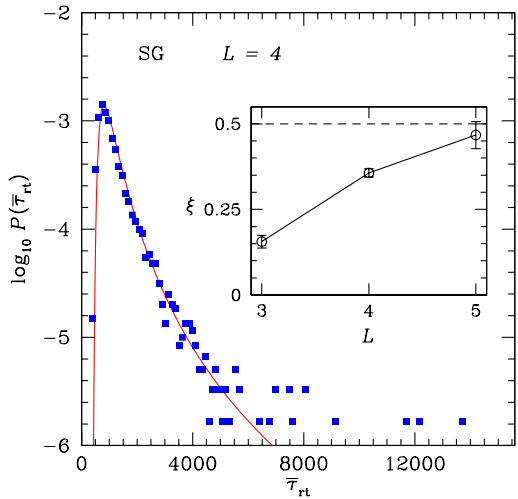}
\vspace*{-1.0cm}

\caption{(Color online)
Distribution of average round-trip times for 5000 different samples of the 
3D Edwards-Anderson Ising spin glass with Gaussian disorder and fixed 
system size $L=4$ in the temperature range from $T_{\rm min} = 0.10$ to
$T_{\rm max} = 2.0$. The data follow a fat-tailed Fr\'echet distribution 
(solid line) with a shape parameter $\xi = 035(1)$. 
The inset shows the shape parameter $\xi$ as a function of system size 
$L$. Already for $L \gtrsim 5$, the shape parameter becomes $\xi \gtrsim 
1/2$, indicating that the variance of the distribution is no longer 
properly defined. The simulations have been performed using a fixed 
temperature set with $M=27$ temperature points distributed such that, on 
average, replica swap moves have a nearly flat acceptance rate.
}
\label{fig:sg_evd-order0}
\end{figure}

For the spin glass there is the additional complexity that different disorder
realizations can lead to strong sample-to-sample variations. Thus one could
also surmise that strong sample-to-sample variations exist for the time it
takes to equilibrate individual samples. This 
becomes evident when measuring the round-trip times for replicas in 
a standard parallel tempering simulation with a fixed temperature set, as 
illustrated in Fig.~\ref{fig:sg_evd-order0} for the Edwards-Anderson Ising
spin glass. We find that the measured round-trip times follow a fat-tailed 
Fr\'echet distribution \cite{gumbel:58,tails} (solid line, fit performed
with R \cite{R}). The integrated generalized extreme value distribution 
is given by:
\begin{equation}
  H_{\xi;\mu;\beta}(\tau) =
  \exp\left[ -\left(1+\xi\frac{\tau-\mu}{\beta}\right)^{1/\xi} \right] \;.
  \label{eq:FrechetDistribution}
\end{equation}
Here $\mu$ represents a generalized most probable value (location 
parameter) and $\beta$ a standard deviation (scale parameter). The value 
of the shape parameter $\xi$ determines whether the distribution is 
thin-tailed ($\xi < 0$, Weibull), Gumbel ($\xi = 0$), or
fat-tailed ($\xi > 0$, Fr\'echet) \cite{gumbel:58}. Note that when
$\xi > 0$, the $m$-th moment of the Fr\'echet distribution exists only
if $|\xi| < 1/m$, e.g., if $\xi > 1/2$ the variance of the distribution is
not properly defined. Our results are in agreement to
similar observations for broad-histogram 
simulations \cite{dayal:04,alder:04,costa:05}. Note that the distribution is
increasingly more fat-tailed as the system size increases (see the inset to
Fig.~\ref{fig:sg_evd-order0}).

The measurement of the round-trip times thus allows to classify individual 
spin-glass samples as ``typical'' with round-trip times in the bulk of the 
distribution or ``hard'' with round-trip times in the tail of the 
distribution. Such a classification can be very useful to shift 
computational resources towards the ``hard'' samples in the course of a 
simulation as these samples potentially might require substantially longer 
simulation times in order to equilibrate. Preliminary tests suggest 
correlations between round-trip and equilibration times.
The observation of strong sample-to-sample 
variations in the distribution of round-trip times also raises the 
general question, whether previous spin-glass studies have  properly 
equilibrated the ``hardest" samples in their simulations.
It remains to be verified whether this introduces a systematic error in 
the analysis of spin-glass systems. Specifically, the finite-size scaling 
should be sensitive to such systematic errors, as it has been observed 
that the number of ``hard'' samples significantly increases with system 
size, see the inset in Fig.~\ref{fig:sg_evd-order0} and 
Refs. \onlinecite{dayal:04}, \onlinecite{alder:04}, and \onlinecite{costa:05}.

On the other hand, we can ask whether we can optimize the simulated 
temperature set and generate a ``common'' temperature set for samples from 
the various parts of the distribution. To do so, we apply the 
feedback optimization outlined above in such a way that we generate a 
common probability distribution ${\bar \eta}(T)$ for a set of samples 
that are each characterized by their own diffusivity $D_i(T)$, 
steady-state fraction $f_i(T)$ and current $j_i$, which are related by
\begin{equation}
  j_i = D_i(T) {\bar \eta}(T) \frac{df_i}{dT},
\end{equation}
where the index $i$ indicates the samples in the given set. 
Rearranging this equation, the local diffusivity of an individual sample 
can be expressed as
\begin{equation}
  D_i(T) = \frac{j_i}{ {\bar \eta}(T) \cdot df_i / dT} \;.
\end{equation}
To ensure equilibration of all samples we want to simulate each sample for a
fixed number of round trips, despite the strong sample-to-sample variations. 
In order to minimize the overall computer time spent to simulate such a set 
of samples, we then minimize the {\em sum} of round-trip times 
$\sum_i \tau_i$. This is equivalent to minimizing the sum of the inverse of
all currents, i.e.,  $j_i^{-1}$, since the current $j_i$ is inversely 
proportional to the round-trip time $\tau_i$. Following a similar line of 
arguments as for the original temperature set optimization, we find that the 
optimal common temperature distribution ${\bar \eta}^{\rm opt}(T)$ is 
proportional to the square root of the {\em sum} of inverse local 
diffusivities
\begin{equation}
  {\bar \eta}^{\rm opt}(T) \propto \sqrt{ \sum_i \frac{1}{D_i(T)} } \;.
\end{equation}
Again we can use a feedback loop to find an optimized common temperature set
by feeding back the measured local diffusivities
\begin{equation}
  {\bar \eta}^{\rm opt}(T) = C \sqrt{ {\bar \eta}(T) \sum_i 
  \left( \tau_i \cdot \frac{df_i}{dT} \right)} \;,
\end{equation}
where $C$ is a normalization constant. The common optimized temperature set is
then found using a partial integration as given in Eq.~(\ref{eq:feedback}).

\section{Conclusions}
\label{sec:conclusions}

We have introduced an approach to systematically optimize temperature sets 
for parallel tempering Monte Carlo simulations using an adaptive feedback 
method that is motivated by a recently developed ensemble optimization 
technique for broad-histogram Monte Carlo simulations \cite{trebst:04}. 
We have applied the method to two paradigmatic spin models, the ferromagnetic 
Ising model and the fully frustrated Ising model in two dimensions. For both 
models we have shown that the feedback technique improves sampling of the
phase space by reducing the average round-trip time of replicas diffusing in 
temperature space. 

Probably our most important result is the insight, that the common wisdom 
that temperature sets in parallel tempering Monte Carlo should yield 
temperature-independent acceptance probabilities for the swap moves is not 
necessarily an optimal choice.
Our feedback algorithm shifts temperature points in the 
optimized temperature sets towards the bottlenecks of the simulation, typically
in the vicinity of a phase transition, where equilibration is suppressed. 
In particular, this has the effect that the acceptance probabilities for 
replica swap moves are higher around the bottlenecks and are {\em not} 
temperature independent for the so-optimized temperature sets.

We also outline an approach to define ``common'' temperature sets for systems
that require configurational averages, such as spin glasses. In addition, we
have briefly discussed the effects of sample-to-sample fluctuations with 
respect to equilibration times of individual spin glass samples, thus 
pointing towards a potential source of systematic errors in previous 
numerical studies of spin glasses.

Clearly a deeper analysis of feedback-optimized parallel tempering Monte
Carlo is needed, in particular in the context of spin glasses, as well
as the questions raised at the end of Sec.~\ref{sec:ff}.

\begin{acknowledgments}
We thank A.~Honecker, S.~D.~Huber, M.~K\"orner, C.~Predescu, K.~Tran, 
D.~W\"urtz and A.~P.~Young for stimulating discussions. 
S.~T.~acknowledges support by the Swiss National Science Foundation.
\end{acknowledgments}

\bibliography{refs,comments}

\begin{thebibliography}{34}
\expandafter\ifx\csname natexlab\endcsname\relax\def\natexlab#1{#1}\fi
\expandafter\ifx\csname bibnamefont\endcsname\relax
  \def\bibnamefont#1{#1}\fi
\expandafter\ifx\csname bibfnamefont\endcsname\relax
  \def\bibfnamefont#1{#1}\fi
\expandafter\ifx\csname citenamefont\endcsname\relax
  \def\citenamefont#1{#1}\fi
\expandafter\ifx\csname url\endcsname\relax
  \def\url#1{\texttt{#1}}\fi
\expandafter\ifx\csname urlprefix\endcsname\relax\def\urlprefix{URL }\fi
\providecommand{\bibinfo}[2]{#2}
\providecommand{\eprint}[2][]{\url{#2}}

\bibitem[{\citenamefont{Trebst et~al.}(2004)\citenamefont{Trebst, Huse, and
  Troyer}}]{trebst:04}
\bibinfo{author}{\bibfnamefont{S.}~\bibnamefont{Trebst}},
  \bibinfo{author}{\bibfnamefont{D.~A.} \bibnamefont{Huse}}, \bibnamefont{and}
  \bibinfo{author}{\bibfnamefont{M.}~\bibnamefont{Troyer}},
  \emph{\bibinfo{title}{{Optimizing the ensemble for equilibration in
  broad-histogram Monte Carlo simulations}}}, \bibinfo{journal}{Phys. Rev. E}
  \textbf{\bibinfo{volume}{70}}, \bibinfo{pages}{046701}
  (\bibinfo{year}{2004}).

\bibitem[{\citenamefont{{Dayal} et~al.}(2004)\citenamefont{{Dayal}, {Trebst},
  {Wessel}, {W{\"u}rtz}, {Troyer}, {Sabhapandit}, and
  {Coppersmith}}}]{dayal:04}
\bibinfo{author}{\bibfnamefont{P.}~\bibnamefont{{Dayal}}},
  \bibinfo{author}{\bibfnamefont{S.}~\bibnamefont{{Trebst}}},
  \bibinfo{author}{\bibfnamefont{S.}~\bibnamefont{{Wessel}}},
  \bibinfo{author}{\bibfnamefont{D.}~\bibnamefont{{W{\"u}rtz}}},
  \bibinfo{author}{\bibfnamefont{M.}~\bibnamefont{{Troyer}}},
  \bibinfo{author}{\bibfnamefont{S.}~\bibnamefont{{Sabhapandit}}},
  \bibnamefont{and} \bibinfo{author}{\bibfnamefont{S.~N.}
  \bibnamefont{{Coppersmith}}}, \emph{\bibinfo{title}{{{Performance Limitations
  of Flat-Histogram Methods}}}}, \bibinfo{journal}{Phys. Rev. Lett.}
  \textbf{\bibinfo{volume}{92}}, \bibinfo{pages}{097201}
  (\bibinfo{year}{2004}).

\bibitem[{\citenamefont{{Berg} and {Neuhaus}}(1991)}]{berg:91}
\bibinfo{author}{\bibfnamefont{B.~A.} \bibnamefont{{Berg}}} \bibnamefont{and}
  \bibinfo{author}{\bibfnamefont{T.}~\bibnamefont{{Neuhaus}}},
  \emph{\bibinfo{title}{{{Multicanonical algorithms for first order phase
  transitions}}}}, \bibinfo{journal}{Phys. Lett. B}
  \textbf{\bibinfo{volume}{267}}, \bibinfo{pages}{249} (\bibinfo{year}{1991}).

\bibitem[{\citenamefont{Berg and Neuhaus}(1992)}]{berg:92}
\bibinfo{author}{\bibfnamefont{B.}~\bibnamefont{Berg}} \bibnamefont{and}
  \bibinfo{author}{\bibfnamefont{T.}~\bibnamefont{Neuhaus}},
  \emph{\bibinfo{title}{Multicanonical ensemble: a new approach to simulate
  first-order phase transitions}}, \bibinfo{journal}{Phys. Rev. Lett.}
  \textbf{\bibinfo{volume}{68}}, \bibinfo{pages}{9} (\bibinfo{year}{1992}).

\bibitem[{\citenamefont{{de Oliveira} et~al.}(1996)\citenamefont{{de Oliveira},
  {Penna}, and {Herrmann}}}]{deoliveira:96}
\bibinfo{author}{\bibfnamefont{P.~M.~C.} \bibnamefont{{de Oliveira}}},
  \bibinfo{author}{\bibfnamefont{T.~J.~P.} \bibnamefont{{Penna}}},
  \bibnamefont{and} \bibinfo{author}{\bibfnamefont{H.~J.}
  \bibnamefont{{Herrmann}}}, \emph{\bibinfo{title}{{{Broad Histogram
  Method}}}}, \bibinfo{journal}{Braz. J. Phys.} \textbf{\bibinfo{volume}{26}},
  \bibinfo{pages}{677} (\bibinfo{year}{1996}).

\bibitem[{\citenamefont{Wang and Swendsen}(2002)}]{wang:02}
\bibinfo{author}{\bibfnamefont{J.-S.} \bibnamefont{Wang}} \bibnamefont{and}
  \bibinfo{author}{\bibfnamefont{R.~H.} \bibnamefont{Swendsen}},
  \emph{\bibinfo{title}{{Transition Matrix Monte Carlo Method}}},
  \bibinfo{journal}{J. Stat. Phys.} \textbf{\bibinfo{volume}{106}},
  \bibinfo{pages}{245} (\bibinfo{year}{2002}).

\bibitem[{\citenamefont{Wang and Landau}(2001{\natexlab{a}})}]{wang:01}
\bibinfo{author}{\bibfnamefont{F.}~\bibnamefont{Wang}} \bibnamefont{and}
  \bibinfo{author}{\bibfnamefont{D.~P.} \bibnamefont{Landau}},
  \emph{\bibinfo{title}{An efficient, multiple-range random walk algorithm to
  calculate the density of states}}, \bibinfo{journal}{Phys. Rev. Lett.}
  \textbf{\bibinfo{volume}{86}}, \bibinfo{pages}{2050}
  (\bibinfo{year}{2001}{\natexlab{a}}).

\bibitem[{\citenamefont{Wang and Landau}(2001{\natexlab{b}})}]{wang:01a}
\bibinfo{author}{\bibfnamefont{F.}~\bibnamefont{Wang}} \bibnamefont{and}
  \bibinfo{author}{\bibfnamefont{D.~P.} \bibnamefont{Landau}},
  \emph{\bibinfo{title}{Determining the density of states for classical
  statistical models: A random walk algorithm to produce a flat histogram}},
  \bibinfo{journal}{Phys. Rev. E} \textbf{\bibinfo{volume}{64}},
  \bibinfo{pages}{056101} (\bibinfo{year}{2001}{\natexlab{b}}).

\bibitem[{\citenamefont{Hukushima and Nemoto}(1996)}]{hukushima:96}
\bibinfo{author}{\bibfnamefont{K.}~\bibnamefont{Hukushima}} \bibnamefont{and}
  \bibinfo{author}{\bibfnamefont{K.}~\bibnamefont{Nemoto}},
  \emph{\bibinfo{title}{Exchange {M}onte {C}arlo method and application to spin
  glass simulations}}, \bibinfo{journal}{J. Phys. Soc. Jpn.}
  \textbf{\bibinfo{volume}{65}}, \bibinfo{pages}{1604} (\bibinfo{year}{1996}).

\bibitem[{\citenamefont{Earl and Deem}(2005)}]{earl:05}
\bibinfo{author}{\bibfnamefont{D.~J.} \bibnamefont{Earl}} \bibnamefont{and}
  \bibinfo{author}{\bibfnamefont{M.~W.} \bibnamefont{Deem}},
  \emph{\bibinfo{title}{{{Parallel Tempering: Theory, Applications, and New
  Perspectives}}}} (\bibinfo{year}{2005}), \bibinfo{note}{(physics/0508111)}.

\bibitem[{\citenamefont{Swendsen and Wang}(1986)}]{swendsen:86}
\bibinfo{author}{\bibfnamefont{R.~H.} \bibnamefont{Swendsen}} \bibnamefont{and}
  \bibinfo{author}{\bibfnamefont{J.}~\bibnamefont{Wang}},
  \emph{\bibinfo{title}{Replica {M}onte {C}arlo simulation of spin-glasses}},
  \bibinfo{journal}{Phys. Rev. Lett.} \textbf{\bibinfo{volume}{57}},
  \bibinfo{pages}{2607} (\bibinfo{year}{1986}).

\bibitem[{\citenamefont{Marinari and Parisi}(1992)}]{marinari:92}
\bibinfo{author}{\bibfnamefont{E.}~\bibnamefont{Marinari}} \bibnamefont{and}
  \bibinfo{author}{\bibfnamefont{G.}~\bibnamefont{Parisi}},
  \emph{\bibinfo{title}{Simulated tempering: A new {M}onte {C}arlo scheme}},
  \bibinfo{journal}{Europhys. Lett.} \textbf{\bibinfo{volume}{19}},
  \bibinfo{pages}{451} (\bibinfo{year}{1992}).

\bibitem[{\citenamefont{{Lyubartsev} et~al.}(1992)\citenamefont{{Lyubartsev},
  {Martsinovski}, {Shevkunov}, and {Vorontsov-Velyaminov}}}]{lyubartsev:92}
\bibinfo{author}{\bibfnamefont{A.~P.} \bibnamefont{{Lyubartsev}}},
  \bibinfo{author}{\bibfnamefont{A.~A.} \bibnamefont{{Martsinovski}}},
  \bibinfo{author}{\bibfnamefont{S.~V.} \bibnamefont{{Shevkunov}}},
  \bibnamefont{and} \bibinfo{author}{\bibfnamefont{P.~N.}
  \bibnamefont{{Vorontsov-Velyaminov}}}, \emph{\bibinfo{title}{{{New approach
  to Monte Carlo calculation of the free energy: Method of expanded
  ensembles}}}}, \bibinfo{journal}{J. Chem. Phys.}
  \textbf{\bibinfo{volume}{96}}, \bibinfo{pages}{1776} (\bibinfo{year}{1992}).

\bibitem[{\citenamefont{Kofke}(2004{\natexlab{a}})}]{kofke:02}
\bibinfo{author}{\bibfnamefont{D.~A.} \bibnamefont{Kofke}},
  \emph{\bibinfo{title}{On the acceptance probability of replica-exchange monte
  carlo trials}}, \bibinfo{journal}{J. Chem. Phys.}
  \textbf{\bibinfo{volume}{117}}, \bibinfo{pages}{6911}
  (\bibinfo{year}{2004}{\natexlab{a}}).

\bibitem[{\citenamefont{Kofke}(2004{\natexlab{b}})}]{kofke:04}
\bibinfo{author}{\bibfnamefont{D.~A.} \bibnamefont{Kofke}},
  \emph{\bibinfo{title}{Comment on "the incomplete beta function law for
  parallel tempering sampling of classical canonical systems" [j. chem. phys.
  120, 4119 (2004)]}}, \bibinfo{journal}{J. Chem. Phys.}
  \textbf{\bibinfo{volume}{121}}, \bibinfo{pages}{1167}
  (\bibinfo{year}{2004}{\natexlab{b}}).

\bibitem[{\citenamefont{{Rathore} et~al.}(2005)\citenamefont{{Rathore},
  {Chopra}, and {de Pablo}}}]{rathore:05}
\bibinfo{author}{\bibfnamefont{N.}~\bibnamefont{{Rathore}}},
  \bibinfo{author}{\bibfnamefont{M.}~\bibnamefont{{Chopra}}}, \bibnamefont{and}
  \bibinfo{author}{\bibfnamefont{J.~J.} \bibnamefont{{de Pablo}}},
  \emph{\bibinfo{title}{{{Optimal allocation of replicas in parallel tempering
  simulations}}}}, \bibinfo{journal}{J. Chem. Phys.}
  \textbf{\bibinfo{volume}{122}}, \bibinfo{pages}{024111}
  (\bibinfo{year}{2005}).

\bibitem[{\citenamefont{Predescu et~al.}(2004)\citenamefont{Predescu, Predescu,
  and Ciobanu}}]{predescu:04}
\bibinfo{author}{\bibfnamefont{C.}~\bibnamefont{Predescu}},
  \bibinfo{author}{\bibfnamefont{M.}~\bibnamefont{Predescu}}, \bibnamefont{and}
  \bibinfo{author}{\bibfnamefont{C.}~\bibnamefont{Ciobanu}},
  \emph{\bibinfo{title}{{The incomplete beta function law for parallel
  tempering sampling of classical canonical systems}}}, \bibinfo{journal}{J.
  Chem. Phys.} \textbf{\bibinfo{volume}{120}}, \bibinfo{pages}{4119}
  (\bibinfo{year}{2004}).

\bibitem[{\citenamefont{Predescu et~al.}(2005)\citenamefont{Predescu, Predescu,
  and Ciobanu}}]{predescu:05}
\bibinfo{author}{\bibfnamefont{C.}~\bibnamefont{Predescu}},
  \bibinfo{author}{\bibfnamefont{M.}~\bibnamefont{Predescu}}, \bibnamefont{and}
  \bibinfo{author}{\bibfnamefont{C.}~\bibnamefont{Ciobanu}},
  \emph{\bibinfo{title}{{On the Efficiency of Exchange in Parallel Tempering
  Monte Carlo Simulations}}}, \bibinfo{journal}{J. Phys. Chem. B}
  \textbf{\bibinfo{volume}{109}}, \bibinfo{pages}{4189} (\bibinfo{year}{2005}).

\bibitem[{\citenamefont{Kone and Kofke}(2005)}]{kone:05}
\bibinfo{author}{\bibfnamefont{A.}~\bibnamefont{Kone}} \bibnamefont{and}
  \bibinfo{author}{\bibfnamefont{D.~A.} \bibnamefont{Kofke}},
  \emph{\bibinfo{title}{{Selection of temperature intervals for
  parallel-tempering simulations}}}, \bibinfo{journal}{J. Chem. Phys.}
  \textbf{\bibinfo{volume}{122}}, \bibinfo{pages}{206101}
  (\bibinfo{year}{2005}).

\bibitem[{\citenamefont{Diep}(2005)}]{diep:05}
\bibinfo{author}{\bibfnamefont{H.~T.} \bibnamefont{Diep}},
  \emph{\bibinfo{title}{{Frustrated Spin Systems}}} (\bibinfo{publisher}{World
  Scientific}, \bibinfo{address}{Singapore}, \bibinfo{year}{2005}).

\bibitem[{\citenamefont{Binder and Young}(1986)}]{binder:86}
\bibinfo{author}{\bibfnamefont{K.}~\bibnamefont{Binder}} \bibnamefont{and}
  \bibinfo{author}{\bibfnamefont{A.~P.} \bibnamefont{Young}},
  \emph{\bibinfo{title}{Spin glasses: Experimental facts, theoretical concepts
  and open questions}}, \bibinfo{journal}{Rev. Mod. Phys.}
  \textbf{\bibinfo{volume}{58}}, \bibinfo{pages}{801} (\bibinfo{year}{1986}).

\bibitem[{\citenamefont{M\'ezard et~al.}(1987)\citenamefont{M\'ezard, Parisi,
  and Virasoro}}]{mezard:87}
\bibinfo{author}{\bibfnamefont{M.}~\bibnamefont{M\'ezard}},
  \bibinfo{author}{\bibfnamefont{G.}~\bibnamefont{Parisi}}, \bibnamefont{and}
  \bibinfo{author}{\bibfnamefont{M.~A.} \bibnamefont{Virasoro}},
  \emph{\bibinfo{title}{Spin Glass Theory and Beyond}}
  (\bibinfo{publisher}{World Scientific}, \bibinfo{address}{Singapore},
  \bibinfo{year}{1987}).

\bibitem[{\citenamefont{Young}(1998)}]{young:98}
\bibinfo{editor}{\bibfnamefont{A.~P.} \bibnamefont{Young}}, ed.,
  \emph{\bibinfo{title}{Spin Glasses and Random Fields}}
  (\bibinfo{publisher}{World Scientific}, \bibinfo{address}{Singapore},
  \bibinfo{year}{1998}).

\bibitem[{\citenamefont{Wales}(2003)}]{wales:03}
\bibinfo{author}{\bibfnamefont{D.~J.} \bibnamefont{Wales}},
  \emph{\bibinfo{title}{{Energy Landscapes}}} (\bibinfo{publisher}{Cambridge
  University Press}, \bibinfo{address}{Cambridge}, \bibinfo{year}{2003}).

\bibitem[{ffi()}]{ffimproblems}
\bibinfo{note}{Due to the discrete energy space for the fully-frustrated Ising
  model the tuning of the temperature points is extremely difficult. Small
  changes to one of the temperature points can change the acceptance
  probabilities drastically. Thus the ``flat'' temperature sets, computed with
  an approximate method due to Predescu (private communication) exhibit
  acceptance probabilities for swap moves which are approximatively constant
  and independent of temperature within 10 -- 20\%.}

\bibitem[{\citenamefont{Edwards and Anderson}(1975)}]{edwards:75}
\bibinfo{author}{\bibfnamefont{S.~F.} \bibnamefont{Edwards}} \bibnamefont{and}
  \bibinfo{author}{\bibfnamefont{P.~W.} \bibnamefont{Anderson}},
  \emph{\bibinfo{title}{Theory of spin glasses}}, \bibinfo{journal}{J. Phys. F:
  Met. Phys.} \textbf{\bibinfo{volume}{5}}, \bibinfo{pages}{965}
  (\bibinfo{year}{1975}).

\bibitem[{\citenamefont{Bhatt and Young}(1988)}]{bhatt:88}
\bibinfo{author}{\bibfnamefont{R.~N.} \bibnamefont{Bhatt}} \bibnamefont{and}
  \bibinfo{author}{\bibfnamefont{A.~P.} \bibnamefont{Young}},
  \emph{\bibinfo{title}{Numerical studies of {I}sing spin glasses in two, three
  and four dimensions}}, \bibinfo{journal}{Phys. Rev. B}
  \textbf{\bibinfo{volume}{37}}, \bibinfo{pages}{5606} (\bibinfo{year}{1988}).

\bibitem[{\citenamefont{Marinari et~al.}(1998)\citenamefont{Marinari, Parisi,
  and Ruiz-Lorenzo}}]{marinari:98}
\bibinfo{author}{\bibfnamefont{E.}~\bibnamefont{Marinari}},
  \bibinfo{author}{\bibfnamefont{G.}~\bibnamefont{Parisi}}, \bibnamefont{and}
  \bibinfo{author}{\bibfnamefont{J.~J.} \bibnamefont{Ruiz-Lorenzo}},
  \emph{\bibinfo{title}{On the phase structure of the 3d {E}dwards {A}nderson
  spin glass}}, \bibinfo{journal}{Phys. Rev. B} \textbf{\bibinfo{volume}{58}},
  \bibinfo{pages}{14852} (\bibinfo{year}{1998}).

\bibitem[{\citenamefont{Katzgraber et~al.}(2006)\citenamefont{Katzgraber,
  K\"orner, and Young}}]{katzgraber:06}
\bibinfo{author}{\bibfnamefont{H.~G.} \bibnamefont{Katzgraber}},
  \bibinfo{author}{\bibfnamefont{M.}~\bibnamefont{K\"orner}}, \bibnamefont{and}
  \bibinfo{author}{\bibfnamefont{A.~P.} \bibnamefont{Young}},
  \emph{\bibinfo{title}{{Detailed study of universality in three-dimensional
  spin glasses}}} (\bibinfo{year}{2006}), \bibinfo{note}{(cond-mat/0602212)}.

\bibitem[{\citenamefont{Gumbel}(1958)}]{gumbel:58}
\bibinfo{author}{\bibfnamefont{E.~J.} \bibnamefont{Gumbel}},
  \emph{\bibinfo{title}{{Statistics of extremes}}}
  (\bibinfo{publisher}{Columbia University Press}, \bibinfo{address}{New York},
  \bibinfo{year}{1958}).

\bibitem[{tai()}]{tails}
\bibinfo{note}{The deviations of the data to the fitting function for large
  average round-trip times can be ascribed to the limited statistics used.}

\bibitem[{R()}]{R}
\bibinfo{note}{R Core Team (R Manuals)},
  \urlprefix\url{http://cran.r-project.org}.

\bibitem[{\citenamefont{{Alder} et~al.}(2004)\citenamefont{{Alder}, {Trebst},
  {Hartmann}, and {Troyer}}}]{alder:04}
\bibinfo{author}{\bibfnamefont{S.}~\bibnamefont{{Alder}}},
  \bibinfo{author}{\bibfnamefont{S.}~\bibnamefont{{Trebst}}},
  \bibinfo{author}{\bibfnamefont{A.~K.} \bibnamefont{{Hartmann}}},
  \bibnamefont{and} \bibinfo{author}{\bibfnamefont{M.}~\bibnamefont{{Troyer}}},
  \emph{\bibinfo{title}{{{Dynamics of the Wang Landau algorithm and complexity
  of rare events for the three-dimensional bimodal Ising spin glass}}}},
  \bibinfo{journal}{J. Stat. Mech.}
  \textbf{\bibinfo{volume}{\normalfont{P07008}}} (\bibinfo{year}{2004}).

\bibitem[{\citenamefont{Costa et~al.}(2005)\citenamefont{Costa, Lopes, and
  Lopes~dos Santos}}]{costa:05}
\bibinfo{author}{\bibfnamefont{M.~D.} \bibnamefont{Costa}},
  \bibinfo{author}{\bibfnamefont{J.~V.} \bibnamefont{Lopes}}, \bibnamefont{and}
  \bibinfo{author}{\bibfnamefont{J.~M.~B.} \bibnamefont{Lopes~dos Santos}},
  \emph{\bibinfo{title}{{Analytical study of tunneling times in flat histogram
  Monte Carlo}}}, \bibinfo{journal}{Europhys. Lett.}
  \textbf{\bibinfo{volume}{72}}, \bibinfo{pages}{802} (\bibinfo{year}{2005}).

\end{thebibliography}

\end{document}